\pdfoutput=1
\RequirePackage{ifpdf}
\ifpdf 
\documentclass[pdftex]{sigma}
\else
\documentclass{sigma}
\fi

\begin{document}

\newcommand{\arXivNumber}{1402.0740}

\allowdisplaybreaks

\renewcommand{\PaperNumber}{087}

\FirstPageHeading

\ShortArticleName{Exact Free Energies of Statistical Systems on Random Networks}

\ArticleName{Exact Free Energies of Statistical Systems\\
on Random Networks}

\Author{Naoki SASAKURA~$^\dag$ and Yuki SATO~$^\ddag$}

\AuthorNameForHeading{N.~Sasakura and Y.~Sato}

\Address{$^\dag$~Yukawa Institute for Theoretical Physics, Kyoto University, Kyoto 606-8502, Japan}
\EmailD{\href{mailto:sasakura@yukawa.kyoto-u.ac.jp}{sasakura@yukawa.kyoto-u.ac.jp}}

\Address{$^\ddag$~National Institute for Theoretical Physics, Department of Physics and Centre\\
\hphantom{$^\ddag$}~for Theoretical Physics, University of the Witwartersrand, WITS 2050, South Africa}
\EmailD{\href{mailto:Yuki.Sato@wits.ac.za}{Yuki.Sato@wits.ac.za}}

\ArticleDates{Received June 10, 2014, in f\/inal form August 07, 2014; Published online August 15, 2014}

\Abstract{Statistical systems on random networks can be formulated in terms of partition functions expressed with
integrals by regarding Feynman diagrams as random networks.
We consider the cases of random networks with bounded but generic degrees of vertices, and show that the free energies
can be exactly evaluated in the thermodynamic limit by the Laplace method, and that the exact expressions can in
principle be obtained by solving polynomial equations for mean f\/ields.
As demonstrations, we apply our method to the ferromagnetic Ising models on random networks.
The free energy of the ferromagnetic Ising model on random networks with trivalent vertices is shown to exactly
reproduce that of the ferromagnetic Ising model on the Bethe lattice.
We also consider the cases with heterogeneity with mixtures of orders of vertices, and derive the known formula of the
Curie temperature.}


\Classification{05C82; 37A60; 46N55; 82B20; 81U15; 83C15}

\section{Introduction}

The Ising model is a~theory of magnets with only nearest-neighbor interactions sitting on a~lattice.
It is a~simple but tremendously useful model in physics which allows us to describe cooperative phenomena at a~critical
point.
Its ef\/fectiveness, in fact, is not exclusive to physics, but expands to various disciplines such as chemistry,
metallurgy, mathematics and biology.
Therefore, needless to say, the Ising model plays a~quite important role in science.

In the history of Ising model, the most inf\/luential work has been made by Onsager in $1944$~\cite{onsager}: the exact
solution of the $2$-dimensional model has been obtained; thereby the existence of the phase transition has been proven.
A~signif\/icance of the exact result is to provide us a~clear-cut picture of the physics near the critical point.
Although $70$ years have gone by since Onsager's novel work, no one has yet solved the $3$-dimensional Ising model
exactly.
Recently, however, there was a~great progress in that direction based on the conformal bootstrap~\cite{ElShowk:2012ht},
although still it has not given us an exact result.
Other examples of the exactly solvable Ising model are, for instance, those on the $2$-dimensional dynamical
lattice~\cite{Kazakov:1986hu} and the Bethe lattice (inf\/inite regular tree)~\cite{Baxter:1982zz}.
In the former, some simplif\/ications exist because of enhanced symmetries.
In the latter, it has been argued in~\cite{johnston} that regular random networks asymptotically approach the Bethe
lattice in the thermodynamic limit.
As can be expected from this fact, some exact results have been obtained for the Ising model on the random
networks~\cite{Bachas:1994qn, dembo,dembo+,Dorogovtsev:2002ix,johnston,leone}, which are in accord with that on the
Bethe lattice.

In this letter, we discuss a~general procedure to obtain exact free energies of statistical systems on random networks
with bounded but generic degrees of vertices by considering Feynman diagrams as random networks.
As demonstrations of our method, we obtain the exact free energy of the ferromagnetic Ising model on random networks
with trivalent vertices, and show that it exactly agrees with that of the Ising model on the Bethe lattice.
We also consider the cases of non-regular random networks with bounded but generic orders of vertices, and derive the
known formula of the Curie point~\cite{Dorogovtsev:2002ix, leone}.
We would like to comment that similar ideas of using Feynman diagrams as random networks were considered in some former
works~\cite{Bachas:1994qn,johnston, whittle}, but the current work has some dif\/ferences in methods and directions of
interests.
For instance, since we stick to exact expressions with real integrations rather than complex ones, the roles played~by
mean f\/ields and the limitation of the mean f\/ield treatment can be clearly seen.

\section{Statistical systems on random networks with trivalent vertices}
Let us start with a~statistical system on random networks with trivalent vertices.
It is def\/ined by the partition function~\cite{ss}
\begin{gather}
Z_n (K,M) =\frac{A}{n!} \int_{-\infty}^\infty \prod\limits^{N}_{a=1} \text{d} \phi_a \left(M_{bcd} \phi_b \phi_c \phi_d
\right)^n e^{- \phi_e K^{-1}_{ef} \phi_f},
\label{z3}
\end{gather}
where $K^{-1}$ is the inverse of a~symmetric real matrix~$K$ with positive eigenvalues, $A=\det{K}^{-\frac{1}{2}}$, $M$~is a~real symmetric three-index tensor, and the repeated indices are assumed to be summed over.
Through the Gaussian integration over $\phi_a$'s, the partition function~\eqref{z3} is represented by the summation of
the Feynman diagrams, which may be regarded as random networks, with~$n$ trivalent vertices weighted by~$M_{abc}$ and
with~$3n/2$ edges weighted by~$K_{ab}$.
The generated networks contain disconnected ones in general, but this possibility can be neglected in the thermodynamic
limit $n\rightarrow \infty$, since a~single connected component dominates in this limit.
By rescaling $\phi_a \to \sqrt{n} \bar{\phi}_a$ in~\eqref{z3} and applying the Laplace method to the integral, one can
obtain the exact free energy in the thermodynamic limit $n \to \infty$~\cite{ss}:
\begin{gather}
f(K,M) = - \lim_{\genfrac{}{}{0pt}{}{n:even}{n\to \infty}} \frac{1}{\beta} \left[\frac{1}{n} \log Z_n (K,M) - f_0
(n)\right] = \frac{1}{\beta} \left(\bar{\phi}_{a} K^{-1}_{ab} \bar{\phi}_{b} - \frac{1}{2} \log \big[g (\bar{\phi})^2
\big] \right),
\label{exactf}
\end{gather}
where~$\beta$ is an inverse temperature, $f_0 (n)=\frac{1}{2} \log n + 1$, $g(\bar \phi)=M_{abc} \bar{\phi}_a
\bar{\phi}_b \bar{\phi}_c$, and $\bar{\phi}_a$ ought to be real and determined in such a~way as to
minimize~\eqref{exactf} for given~$K$,~$M$.
Here we have subtrac\-ted~$f_0(n)$, which is divergent in the limit $n\rightarrow \infty$, since it is independent of~$K$
and~$M$ and hence can be regarded as the free energy of the network rather than that of the statistical system on it.
In taking the limit, we have to assume~$n$ to be even, since $Z_{n={\rm odd}}=0$ for trivalent vertices.
We stress here that the free energy in the thermodynamic limit~\eqref{exactf} is exact, since the corrections to the
Laplace method are of order~$\log [n]/n$ and vanish in the limit.

The minimization of~\eqref{exactf} can be realized by one of the solutions to the following extremeness condition
\begin{gather}
2K^{-1}_{ab} \bar{\phi}_{b} - \frac{3M_{abc} \bar{\phi}_b \bar{\phi}_c}{g(\bar \phi)} =0.
\label{external}
\end{gather}
In order to analyze the condition~\eqref{external}, it is useful to rescale $\bar{\phi}_a = g(\bar \phi) w_a$.
Then~\eqref{external} becomes
\begin{gather}
2K^{-1}_{ab} w_{b} - 3 M_{abc} w_b w_c =0,
\label{eq:w}
\end{gather}
and the free energy turns out to be
\begin{gather}
f(K,M) = \frac{1}{\beta} \left(\frac{3}{2} - \frac{1}{2} \log \left[\frac{3}{2} \right] + \frac{1}{2} \log \big[w_a
K^{-1}_{ab} w_b \big] \right).
\label{eq:fKM}
\end{gather}
Thus the free energy is determined by one of the non-vanishing solutions to~\eqref{eq:w} which minimizes~\eqref{eq:fKM}.

The ferromagnetic Ising model with spins on vertices on random networks can be realized by taking $N=2$ and the weights
as
\begin{gather}
K_{ab} = \exp \left[\beta J \sigma_a \sigma_b\right], \qquad M_{abc}=\exp[\beta H \sigma_a] \delta_{ab}\delta_{ac},
\label{ising}
\end{gather}
where $\sigma_1=1$, $\sigma_2=-1$, and~$H$ and~$J$ are a~magnetic f\/ield and a~nearest-neighbour coupling, respectively.
Here we are forced to consider the ferromagnetic case, $J>0$, since all the eigenvalues of~$K$ must be positive for the
integration~\eqref{z3} to be well-def\/ined.
Then~\eqref{eq:w} are given~by
\begin{gather}
3e^{\beta H} \sinh \left[2 \beta J \right] w^{2}_{1} - e^{\beta J} w_1 + e^{- \beta J} w_2 =0,
\label{w1}
\\
3e^{- \beta H} \sinh \left[2 \beta J \right]w^{2}_{2} - e^{\beta J}w_2 + e^{- \beta J} w_1 =0.
\label{w2}
\end{gather} 
Note that~\eqref{w1} and~\eqref{w2} are symmetric under the simultaneous f\/lip, $H\rightarrow -H$ and $w_1\leftrightarrow
w_2$.
By solving~\eqref{w1} for $w_2$, and plugging it into~\eqref{w2}, one obtains
\begin{gather}
w_1\big[2-6 e^{2 \beta J} \cosh[\beta(H - J)]w_1 + 18 e^{2 \beta J} \sinh[2 \beta J]w_1^2 -27 e^{\beta(H + J)} \sinh[2
\beta J]^2 w_1^3\big]=0.
\label{eq:w1three}
\end{gather}
Since the solutions of~\eqref{eq:w1three} are given by $w_1=0$ and those to a~cubic equation, there exist explicit
analytic expressions for the solutions.
While these solutions are too complicated to write down in the general case, the non-vanishing solutions in the case of
$H=0$ have the following simple expressions:
\begin{alignat*}{3}
& (\hbox{I})\quad &&  w_1=w_2=\frac{1}{3\cosh[\beta]}, &
\\
& (\hbox{II})\quad && w_1=\frac{1\pm\sqrt{\frac{e^{\beta}-3 e^{- \beta}}{2 \cosh[\beta]}}}{6 \sinh[\beta]},\qquad
w_2=\frac{1\mp\sqrt{\frac{e^{\beta}-3 e^{-\beta}}{2 \cosh[\beta]}}}{6 \sinh[\beta]},&
\end{alignat*}
where we have set $J=1$ for simplicity.
The solution (I) is real for all~$\beta$.
The solutions (II) are real and minimize~\eqref{eq:fKM} for $\beta > \beta_c= \frac{1}{2} \log[3]$, but are not real and
therefore inappropriate for $\beta<\beta_c$.
Thus, by putting the solutions into~\eqref{eq:fKM}, the free energy is determined to be

for $\beta \le \beta_c$,
\begin{gather*}
f(\beta) = \frac{3}{2\beta} - \frac{1}{2 \beta} \log\biggl[\frac{3}{2} \biggl] - \frac{1}{2\beta} \log \left[9
\cosh[\beta]^3 \right],
\end{gather*}

for $\beta > \beta_c$,
\begin{gather*}
f(\beta) = \frac{3}{2\beta} - \frac{1}{2\beta} \log\biggl[\frac{3}{2} \biggl] - \frac{1}{2\beta}\log \left[\frac{36
\cosh[\beta]\sinh[\beta]^3}{e^{\beta}-2 e^{-\beta}} \right].
\end{gather*}
From these expressions of the free energy, one can derive the specif\/ic heat, $C = -2 \beta^2 \frac{\partial}{\partial
\beta} f- \beta^3 \frac{\partial^2}{\partial \beta^2} f$:
\begin{gather*}
C(\beta)=
\begin{cases} \dfrac{3\beta^2}{2\cosh[\beta]^2}, & \hbox{for} \ \beta < \beta_c,
\vspace{1mm}\\
\dfrac{3 \beta^2 (2 \sinh [2\beta] + \cosh[2\beta] -2 )}{2\cosh[\beta]^2 \sinh[\beta]^2 (2 - e^{2\beta})^2},
&\hbox{for} \ \beta > \beta_c.
\end{cases}
\end{gather*}

In the case with a~non-vanishing magnetic f\/ield, it is a~straightforward task to obtain the perturbative solutions,
$w_1=w_1^{(0)}+w_1^{(1)} H+ w_1^{(2)} H^2+\cdots$, of~\eqref{eq:w1three}, and compute the free energy in perturbation
of~$H$.
Then one can determine the magnetization $m=-\frac{\partial f}{\partial H}|_{H=0}$ and the magnetic susceptibility
$\chi=-\frac{\partial^2 f}{\partial H^2}|_{H=0}$ as follows:
\begin{gather*}
m(\beta)=
\begin{cases} 0, & \hbox{for} \ \beta < \beta_c,
\\
\dfrac{1}{1-2e^{-2\beta}}\sqrt{\dfrac{e^{2\beta} -3}{e^{2\beta} +1}}, &\hbox{for}\ \beta > \beta_c,
\end{cases}
\\
\chi(\beta)=
\begin{cases} \dfrac{2\beta}{3e^{-2\beta} -1},& \hbox{for} \ \beta < \beta_c,
\vspace{1mm}\\
\dfrac{4 \beta e^{2\beta}}{(2 - e^{2\beta})^2 (e^{4\beta} -2e^{2\beta} -3)},&\hbox{for} \ \beta > \beta_c.
\end{cases}
\end{gather*}

The results above show that there is a~second-order phase transition point at $\beta=\beta_c$ with mean f\/ield features;
the specif\/ic heat has a~jump at the point; the magnetization and the magnetic susceptibility behave around it as $m
(\beta) \sim (\beta-\beta_c)^{1/2}$ and $\chi (\beta) \sim (\beta - \beta_c)^{-1}$, respectively~\cite{Bachas:1994qn, dembo,dembo+,Dorogovtsev:2002ix,johnston,leone}.
In our method, the roles of mean f\/ields are played by $\bar \phi_a$.

In fact, one can show that the system is fully equivalent to the ferromagnetic Ising model on the Bethe lattice of
trivalent vertices.
Let us denote $x=w_2/w_1$.
Then, from~\eqref{w1} and~\eqref{w2}, one can show
\begin{gather}
x=\frac{e^{\beta (-J+H)}+e^{\beta (J-H)} x^2}{e^{\beta(J+H)}+e^{\beta(-J-H)} x^2}.
\label{eq:recur}
\end{gather}
This is the f\/ixed point equation of the recursive relation for solving the ferromagnetic Ising model on the Bethe
lattice, and has been previously derived from saddle point analysis for $H=0$ in the approaches of~\cite{Bachas:1994qn,
johnston}.
From~\eqref{w1}, one can also show
\begin{gather}
w_1=\frac{e^{-\beta(H + J)} \big(e^{2\beta J} - x\big)}{3 \sinh[2 \beta J]}.
\label{eq:w1exp}
\end{gather}
Then, by using~\eqref{eq:recur} and~\eqref{eq:w1exp} to express~$H$ and $w_i$ in terms of~$J$ and~$x$, one obtains
\begin{gather*}
w_a K^{-1}_{ab} w_b= \frac{4 e^{-3 \beta J} \big(z^2 + 1 - z \big(x + x^{-1}\big)\big) \big(x + x^{-1} - 2 z\big)}{9\big(1 - z^2\big)^3},
\end{gather*}
where $z=e^{-2\beta J}$.
By putting this expression into~\eqref{eq:fKM}, one reproduces the free energy of the ferromagnetic Ising model on the
Bethe lattice~\cite{Baxter:1982zz} up to $\text{const}/\beta$, which is irrelevant to the thermodynamic properties.

The above discussions so far for the Ising model are restricted to the ferromagnetic case, $J>0$, and below let us
comment on the anti-ferromagnetic case, $J<0$, in view of our framework.
The restriction comes from the fact that, for $J\le 0$,~$K$ in~\eqref{ising} has a~vanishing or negative eigenvalue
and~\eqref{z3} is ill-def\/ined.
So, to change the form of~$K$, let us consider a~transformation, $K'_{a'b'}=K_{ab}R^{-1}_{aa'} R^{-1}_{bb'} $,
$M'_{a'b'c'}=R_{a'a}R_{b'b}R_{c'c} M_{abc}$, with a~matrix~$R$, which is generally allowed to be {\it complex}.
Obviously, this transformation does not change the statistical weights of each Feynman diagram, and therefore the
statistical systems before and after the transformation can be regarded identical.
Indeed it is easy to explicitly f\/ind an~$R$, by which~$K$ and~$M$ for $J<0$ in~\eqref{ising} are transformed to
a~symmetric real matrix $K'$ with positive eigenvalues and $M'$ which is {\it complex} for $H\neq 0$.
Then the expression~\eqref{z3} with $K'$ and $M'$ is well-def\/ined, and gives the exact free energy for the
anti-ferromagnetic case.
In this case, however, we cannot apply the Laplace method to obtain~\eqref{exactf}, because the integrand is complex.
As a~result, the system cannot be treated with mean f\/ields.
On the other hand, the integration in~\eqref{z3} for such a~case could be computed by the saddle point method by
deforming the real integration contours to complex ones.
However, it would be a~non-trivial issue to generally show the existence of such deformed complex contours satisfying
the conditions validating saddle point treatment, while this would be possible for some concrete cases~\cite{Bachas:1994qn,johnston}.
We would also like to comment that, by considering the tensor products of a~number of~$K$ and~$M$~\cite{Bachas:1994qn},
the replica trick to quench random networks can also be treated in our method.

\section{Extention to random networks with bounded\\ but generic degrees of vertices}

From now on, we will consider random networks with bounded but generic degrees of vertices.
The partition function of such a~generic model is def\/ined~by
\begin{gather}
Z_{n}(K,M,t)=\frac{A}{n!} \int_{-\infty}^\infty \prod\limits_{a=1}^N \text{d} \phi_a \ [g_{n}(\phi)]^n \, e^{-\phi_b
K^{-1}_{bc}\phi_c},
\label{gz}
\end{gather}
where $g_{n}(\phi)$ is a~polynomial function of $\phi_a$ in the form
\begin{gather}
g_{n}(\phi)= \sum\limits^{p}_{k=3} n^{-\frac{k-3}{2}} t^{(k)} M^{(k)}_{a_1 \dots a_k}\phi_{a_1} \cdots \phi_{a_k}.
\label{eq:ggeneral}
\end{gather}
Here $t^{(k)}$ are variables introduced for convenience, and the scalings in~$n$ are needed for all the couplings
$M^{(k)}$ to play roles in dynamics, i.e., without the scalings, the dynamics will be dominated by the highest
order term.
This will become clear in due course.

The generic random networks generated by~\eqref{gz} and~\eqref{eq:ggeneral} contain $3$- to~$p$-valent vertices.
Since the number of vertices with degree~$k$ in each contribution to the partition function can be counted by the order
of $t^{(k)}$, the statistical average of vertex degree can be computed~by
\begin{gather}
\langle k \rangle_n = \frac{1}{n}\sum\limits^{p}_{k=3} k \, t^{(k)}\frac{\partial}{\partial t^{(k)}} \log Z_n (K,M,t).
\label{eq:avk}
\end{gather}

As for the thermodynamic limit, the same rescaling of $\phi_a$ and the application of the Laplace method as before
in~\eqref{gz} leads to that the free energy $f(K,M,t)$ is given by the same expression as~\eqref{exactf} with
\begin{gather*}
g(\bar \phi)= \sum\limits^{p}_{k=3} t^{(k)} M^{(k)}_{a_1 \dots a_k}\bar \phi_{a_1} \cdots \bar \phi_{a_k}.
\end{gather*}
Then the average of vertex degree~\eqref{eq:avk} can be expressed in the thermodynamic limit as
\begin{gather}
\langle k \rangle = - \beta \sum\limits^{p}_{k=3} k\,t^{(k)}\frac{\partial}{\partial t^{(k)}} f(K,M,t)
=\frac{1}{g(\bar\phi)} \bar \phi_a \frac{\partial}{\partial \bar \phi_a}g(\bar\phi).
\label{eq:expressk}
\end{gather}
Similarly, the average of vertex degree square is given~by
\begin{gather}
\langle k^2 \rangle=\frac{1}{g(\bar\phi)}\left(\bar \phi_a \frac{\partial}{\partial \bar \phi_a}\right)^2 g(\bar\phi).
\label{eq:expressk2}
\end{gather}

Let us consider the ferromagnetic Ising model on the general random networks for a~vanishing magnetic f\/ield $H=0$~by
taking $N=2$ and
\begin{gather*}
K_{ab}= \exp [\beta J \sigma_a \sigma_b  ],
\qquad
M^{(k)}_{a_1a_2\ldots a_k}=\delta_{a_1a_2}\delta_{a_1a_3}\cdots \delta_{a_1a_k}.
\end{gather*}
Hereafter we set $J=1$ for simplicity.
The variables $t^{(k)}$ parametrize the weights of each kind of vertices.

The extremeness condition for $\bar \phi_a$ is given by
\begin{gather}
2 K^{-1}_{ab}\bar \phi_b -\frac{\bar g'(\bar \phi_a)}{\bar g(\bar \phi_1) +\bar g(\bar \phi_2)}=0,
\label{eq:extreme}
\end{gather}
where we have introduced $ \bar g(\bar \phi)=\sum\limits_{k=3}^p t^{(k)} {\bar \phi}^k, $ and its derivative $\bar
g'(\bar \phi)$ with respect to $\bar \phi$.
In the disordered phase, the solution is symmetric under $1\leftrightarrow 2$, i.e., $\bar \phi_1=\bar \phi_2=\bar
\phi$.
Therefore $\bar \phi$ must satisfy
\begin{gather}
4 \big(K^{-1}_{11}+K^{-1}_{12}\big)\bar \phi^2 -\langle k \rangle=0,
\label{eq:phikrel}
\end{gather}
which is obtained by substituting $\bar \phi_1=\bar \phi_2=\bar \phi$ into~\eqref{eq:extreme}, and
using~\eqref{eq:expressk} and $K_{11}=K_{22}$, $K_{12}=K_{21}$.
At a~second-order critical point,~\eqref{eq:extreme} must have degenerate solutions.
This requires that the derivatives of~\eqref{eq:extreme} for $a=1,2$,
\begin{gather*}
2 K^{-1}_{ab} \text{d} \bar \phi_b -\frac{\bar g''(\bar \phi_a) \text{d} \phi_a}{\bar g(\bar \phi_1) +\bar g(\bar
\phi_2)} +\frac{\bar g'(\bar \phi_a) \bar g'(\bar \phi_b) \text{d} \bar \phi_b}{(\bar g(\bar \phi_1) +\bar g(\bar
\phi_2))^2},
\end{gather*}
must degenerate at the critical point.
By taking the subtraction and addition of the two derivatives for $a=1,2$, and taking into account $\bar \phi_1=\bar
\phi_2=\bar \phi$,~\eqref{eq:expressk} and~\eqref{eq:expressk2}, one can f\/ind that there exist two cases of degeneracy:
\begin{alignat*}{3}
& \hbox{(A)}\quad && 4 \big(K_{11}^{-1}-K_{12}^{-1}\big) \bar \phi^2-\langle k(k-1) \rangle=0,  &
\\
& \hbox{(B)}\quad && 4 \big(K_{11}^{-1}+K_{12}^{-1}\big)\bar \phi^2-\langle k(k-1) \rangle+\langle k \rangle^2=0.
\end{alignat*}
In the case (A), by using~\eqref{eq:phikrel}, one can delete $\bar \phi^2$, and obtain
\begin{gather*}
\beta_c=\frac{1}{2} \log \frac{\langle k^2 \rangle}{\langle k^2 \rangle -2 \langle k \rangle}.
\end{gather*}
This agrees with the formula for the Curie temperature obtained in the literatures~\cite{Dorogovtsev:2002ix, leone}.
On the other hand, for the case (B), we obtain
\begin{gather*}
\langle k^2 \rangle -\langle k \rangle^2 -2 \langle k \rangle=0.
\end{gather*}

The degeneracy condition of the case (A) is obtained by the subtraction of the two derivatives, and correspondingly the
degeneracy is in the direction $\text{d}\phi_1-\text{d}\phi_2$.
This direction is odd under the exchange $1\leftrightarrow 2$, which is associated to the f\/lip of the magnetic f\/ield,
$H\rightarrow -H$, as explained in the sentence below~\eqref{w2}.
Thus the case (A) corresponds to a~critical point of spontaneous magnetization.
On the other hand, in the case (B), the degeneracy is in the symmetric direction $\text{d}\phi_1+\text{d}\phi_2$, and
the corresponding critical point, if it existed, would not be related to spontaneous magnetization.

\section{Conclusions}

We summarize our results.
To begin with, we have formulated a~systematic general method to obtain the exact free energies in the thermodynamic
limit of statistical systems on random networks with bounded but generic degrees of vertices.
Then, as demonstrations, we have discussed the ferromagnetic Ising models on random networks, and have reproduced the
former results in the literatures~\cite{Bachas:1994qn,dembo,dembo+,Dorogovtsev:2002ix,johnston, leone}.
The mean f\/ield results in the former study can be naturally understood by the roles of~$\bar \phi_a$ in the current
work.
The anti-ferromagnetic case has been shown to have a~dif\/f\/iculty in the mean-f\/ield treatment.
As for the future study, we can in principle put and analyze various statistical systems on random networks with
heterogeneity of degrees of vertices by considering various choices of~$N$,~$K$,~$M$.
It seems also possible to extend the method to include networks with unbounded degrees of vertices by considering
non-polynomial~$g_n(\phi)$.
It would also be interesting to pursue the connection to a~model of quantum gravity~\cite{ss}.

\subsection*{Acknowledgements}

We would like to thank Des Johnston for some communications.

\pdfbookmark[1]{References}{ref}
\LastPageEnding

\end{document}